\documentclass[a4paper]{article}
\usepackage[margin=3.6cm]{geometry}
\usepackage{amssymb}
\usepackage{amsmath,amssymb}
\usepackage{graphicx}
\usepackage{subfigure}

\newtheorem{theorem}{Theorem}

\newtheorem{remark}[theorem]{Remark}
\newenvironment{proof}[1][Proof]{\textbf{#1.} }

\begin{document}

\title{A multistep algorithm for ODEs}

\author{ Marius-F. Danca\\
Romanian Institute of Science and Technology\\
Str. Ciresilor nr. 29, 400487 Cluj-Napoca, Romania \\
danca@rist.ro\\
400114, Cluj-Napoca, Romania}

\maketitle

\begin{abstract} The objective of this paper is to prove the
convergence of a linear implicit multi-step numerical method for
ordinary differential equations. The algorithm is obtained via
Taylor approximations. The convergence is proved following the
Dahlquist theory. As an additional topic, the time stability is
established too. Comparative tests between some of the most known
numerical methods and this method are presented.
\end{abstract}


\noindent\textbf{AMS (MOS)} subject classification: 65L06, 65P99.

\section{Introduction}

\indent

We present a linear implicit \emph{m-}step method LIL (Local
Iterative Linearization) and prove its convergence applied for the
following initial value problem
\begin{equation}
\overset{.}{x}=f\,(t,x),\quad x(t_{0})=x_{0},  \label{1.1}
\end{equation}

\noindent \noindent where \ $f\,:\,[t_{0},T]\times \mathbb{R}^{n}\rightarrow
\mathbb{R}^{n}$,\ $T>0,\,t_{0}\in \mathbb{R}_{+},$\ is a $C^{\,m}$ smooth
Lipschitz function\footnote{%
The Lipschitz condition is necessary for the stability proof.}.

Although the classical linear multi-step algorithms are very known and
utilized, the LIL characteristics (convergence properties, time stability
and applications results) show that this numerical method could be
considered as an interesting alternative to the widely used formulas.

The backward approximation of derivatives implies null coefficients of the
odd order derivatives which represent a major advantage for the propagation
of errors.

As a comparative test two simple ODEs with known analytical solutions and a
chaotic continuous-time dynamical system, first studied by Fabrikant and
Rabinovich [6] and recent numerically re-examined by Danca and Chen [3], was
integrated using the LIL algorithm and some of the most known algorithms.
The complex dynamic of this special model represented a real challenge for
almost all of these methods as shown in Sect.5.

Being an implicit method, an extrapolation is used as the predictor phase.
Like all the $m$-step algorithms, the previous \ $m$ \ points (beside the
first \ $m$ \ start points) should be estimated every step.

To study the convergence we use the unified approach of stability and
consistency developed by Germund Dahlquist in 1956 [2] (see also [7-8]).
Thus, the LIL method applied to the initial value problem (\ref{1.1}) is
considered convergent if and only if it is stable and consistent.

The content of this paper is as follows: In Sect. 2 the LIL method is
deduced. The convergence is proved in Sect. 3. In Sect. 4 is presented the
time stability with the corresponding time stability domains. In Sect. 5
three examples are presented. All computer tests were realized using a Turbo
Pascal code written by the author. In the Appendix, the coefficients of LIL
method are presented.

\section{Deduction of the LIL method}

\indent

Let us consider the uniform grid
\begin{equation*}
\Delta =(t_{0}<t_{1}<...<t_{n}=T\,),\quad n\in \mathbb{N}^{\ast },
\end{equation*}
\noindent with the step-size \
\begin{equation*}
h=\frac{T-t_{0}}{n}=2\,\delta t\,,
\end{equation*}
\strut where \ $\delta t$ \ stands for the ray of the neighborhood \ $%
V_{k}=\left( t_{k}-\delta t\,,\,t_{k}+\delta t\right) \,,$ \ $k=1,2,...,n-1.$

\noindent We assume that all infinite Taylor series converge, but this is
not necessarily since one truncate at a sufficiently large but finite number
of terms.

\noindent One introduce the following notations
\begin{eqnarray*}
x_{k-j} &:&=x(t_{k-j})=x\left[ t_{0}+\left( k-j)h\right) \right] , \\
x_{k}^{(i)} &:&=x^{(i)}(t_{k}),\quad x_{k}^{(0)}:=x(t_{k}),\quad \qquad
j=1,2,...,m.
\end{eqnarray*}
In the following \ $k$ \thinspace is supposed to take the values \ $%
k=1,2,...,n-1.$

\noindent If we consider \ $x_{k-j}$ \ as a function of variable $h$ \
defined in \ $V_{k}$ , then the first \ $m$ \ terms of Taylor approximation
of \ $x_{k-j}$ \ is\footnote{%
The choice of \ $m$ \ and \ $h$ \ is supposed to be such that the Taylor
approximation can be used. The link between \ $h$ \ and \ $m$ \ is analyzed
in Section 3.1}
\begin{equation}
x_{k-j}\thickapprox x_{k}-\frac{j\,h}{1!}x_{k}^{^{\prime }}+\frac{\left(
j\,h\right) ^{2}}{2!}x_{k}^{^{\prime \prime }}-...+\left( -1\right) ^{m}%
\frac{\left( j\,h\right) ^{m}}{m!}x_{k}^{\left( m\right) },
\label{2.1}
\end{equation}
where \ $j=1,2,...m.$ \ The relations \ (\ref{2.1}) represent a Cramer
system with the unknown \ $x_{k}^{\left( i\right) },\ i=1,2,...,m$%
\begin{equation}
\begin{array}[t]{l}
x_{k-1}-x_{k}\thickapprox \frac{h}{1!}x_{k}^{^{\prime }}+\frac{\,h^{2}}{2!}%
x_{k}^{^{\prime \prime }}-...+\left( -1\right) ^{m}\frac{\,h^{m}}{m!}%
x_{k}^{\left( m\right) }, \\
x_{k-2}-x_{k}\thickapprox \frac{2\,h}{1!}x_{k}^{^{\prime }}+\frac{\left(
2\,h\right) ^{2}}{2!}x_{k}^{^{\prime \prime }}-...+\left( -1\right) ^{m}%
\frac{\left( 2\,h\right) ^{m}}{m!}x_{k}^{\left( m\right) }, \\
... \\
x_{k-m}-x_{k}\thickapprox \frac{m\,h}{1!}x_{k}^{^{\prime }}+\frac{\left(
m\,h\right) ^{2}}{2!}x_{k}^{^{\prime \prime }}-...+\left( -1\right) ^{m}%
\frac{\left( m\,h\right) ^{m}}{m!}x_{k}^{\left( m\right) }.
\end{array}
\label{2.2}
\end{equation}
The determinant of the system (\ref{2.2}) is
\begin{equation*}
\Delta =\frac{h^{^{m\left( m+1\right) /2}}}{1!2!...m!}\left|
\begin{array}{lllll}
1 & 1 & 1 & ... & 1 \\
2 & 2^{2} & 2^{3} & ... & 2^{m} \\
... & ... & ... & ... & ... \\
m & m^{2} & m^{3} & ... & m^{m}
\end{array}
\right| =h^{^{^{m\left( m+1\right) /2}}}1!2!...\left( m-2\right) !\,.
\end{equation*}
Because for \ $m\geq 2$\ $\,$we have $\,\Delta \neq 0$,\ \ there exists a
unique solution \vspace{-0.2cm}
\begin{eqnarray*}
x_{k}^{\left( i\right) } &=&\frac{1}{h^{i}}\sum_{j=0}^{m}\delta
_{i\,j}x_{k-j},\quad i=2,...,m\quad \text{for \ \ }m\geq i>1,  
\label{2.3} \\
x_{k}^{^{\prime }} &=&\frac{x_{k}-x_{k-1}}{h}\quad \text{for \ }m=i=1,
\notag
\end{eqnarray*}
the coefficients \ $\delta _{i\,j}$ \ being drawn in Table
7/Appendix.

\noindent Thus we obtained a backward approximation of derivatives, which
represents the key of LIL method. The Taylor approximation of the solution \
$x$, considered now as function of \ $t$ \ in the neighborhood \ $V_{k},$ \
is

\begin{equation}
x(t)\thickapprox x(t_{k})+\frac{t-t_{k}}{1!}x^{^{\prime }}(t_{k})+\frac{%
\left( t-t_{k}\right) ^{2}}{2!}x^{^{\prime \prime
}}(t_{k})+...+\frac{\left( t-t_{k}\right) ^{m}}{m!}x^{\left(
m\right) }(t_{k}).  \label{2.4}
\end{equation}
Next, integrating (\ref{2.4}) in $\ V_{k}$ \ we get \vspace{-0.5cm}

\begin{center}
\begin{equation}
\begin{array}{l}
\int\limits_{t_{k}-\delta \,t}^{t_{k}+\delta
\,t}x(t)\,dt=\int\limits_{-\,\delta \,t}^{\delta
\,t}x(t+t_{k})\,dt\thickapprox \int\limits_{-\,\delta \,t}^{\delta
\,t}\left( \sum\limits_{i\,=0}^{m}x_{k}^{\left( i\right) }t^{i}\right) dt=
\\
=\sum\limits_{\substack{ i\,=0,2,4,...  \\ i\leq \,m}}^{m}\frac{1}{%
2^{i}\left( i+1\right) !}h^{i+1}x_{k}^{\left( i\right) }=h\,x_{k}+h^{3}\frac{%
1}{24}x_{k}^{^{\prime \prime }}+h^{5}\frac{1}{1920}x_{k}^{\left( 4\right)
}+...
\end{array}
\label{2.5}
\end{equation}
\end{center}

\begin{remark}
The zero coefficients of the derivatives \ $x_{k}^{\left( 2i+1\right) }$ \
for $\,i=0,1,2,...$ in\emph{\ (\ref{2.5})} represent a major advantage for
the propagation of errors and computation time.
\end{remark}

\noindent If we use in (\ref{2.5}) the derivatives expression (2.3) we have
\begin{equation}
\int\limits_{t_{k}-\delta \,t}^{t_{k}+\delta
\,t}x(t)\,dt\thickapprox h\sum_{i=0}^{m}\sigma _{0\,i}x_{k-i}\,
\label{2.6}
\end{equation}
the coefficients \ $\sigma _{0\,i}$ \ being given in Table
8(a)/Appendix. Using the same way one can approximate \ $x^{\prime
}$ \ on \ $V_{k}$
\begin{equation}
\int\limits_{t_{k}-\delta \,t}^{t_{k}+\delta \,t}x^{\prime
}(t)\,dt\thickapprox \sum_{i=0}^{m}\sigma _{1\,i}x_{k-i}\,,
\label{2.7}
\end{equation}
the coefficients \ $\sigma _{1i}$ \ being drawn in Table
8(b)/Appendix.

\noindent To overcome the difficulty of Taylor approximation of the
composite function \ $f$\ \ we found, empirically, that the relations (\ref
{2.6}) could be considered as a simple way to approximate the integral of \ $%
f$ \ without altering the method convergence. Thus
\begin{equation}
\int\limits_{t_{k}-\delta \,t}^{t_{k}+\delta
\,t}f\,(t,x(t))\,dt\thickapprox h\sum_{i\,=\,0}^{m}\sigma
_{0\,i}\,\,f_{k-i},  \label{2.8}
\end{equation}
where \ $f_{k-i}:=f\,(t_{k-i},x(t_{k-i}))$ .

\noindent Using (\ref{2.7}) and (\ref{2.8}) we can integrate (\ref{1.1}) in
\ $V_{k}\,$
\begin{equation*}
\sum_{i\,=\,0}^{m}\sigma _{1i}x_{k-i}=h\,\sum_{i\,=\,0}^{m}\sigma
_{0\,i}\,\,f_{k-i}\,.
\end{equation*}
\noindent Because \ $\sigma _{10}\neq 0$ , \ for every \ $m\,\ $(see
Table 8/Appendix), \ the approximation of the solution in \ $V_{k}$
\ is
\begin{equation}
x_{k}=\frac{\,h}{\sigma _{10}}\sum\limits_{i\,=\,0}^{m}\sigma
_{0\,i}\,\,f_{k-i}-\frac{1}{\sigma
_{10}}\sum\limits_{i\,=\,1}^{m}\sigma _{1\,i}\,x_{k-i}.  \label{2.9}
\end{equation}
\noindent If we denote
\begin{equation*}
u_{k}:=\frac{\,1}{\sigma _{10}}\sum\limits_{i\,=\,0}^{m}\sigma
_{0\,i}\,\,f_{k-i},\quad v_{k}:=-\,\frac{1}{\sigma _{10}}\sum\limits_{i\,=%
\,1}^{m}\sigma _{1\,i}\,x_{k-i},
\end{equation*}
the relations (\ref{2.9}) become
\begin{equation}
x_{k}=v_{k}+h\,u_{k}\,,\quad k=1,2,...n-1\,.  \label{2.10}
\end{equation}
Formula (\ref{2.10}) represents the $m\,$th-order LIL method. In
Table 1 the formulae for orders one through five ($m\in
\{1,2,3,4,5\})$ are presented.

\label{Table 1}
\begin{table}
\begin{center}
\begin{tabular}{||l|l||}
\hline\hline
$m$ &  \\ \hline
&  \\
$1$ & $x_{k}=x_{k-1}+h\,f_{k},$ \\
&  \\ \hline
&  \\
$2$ & $x_{k}=\frac{4}{3}x_{k-1}-\frac{1}{3}x_{k-2}+\frac{h}{36}\left(
25\,f_{k}-2\,f_{k-1}+f_{k-2}\right) ,$ \\
&  \\ \hline
&  \\
$3$ & $x_{k}=\frac{5}{3}x_{k-1}-\frac{13}{15}x_{k-2}+\frac{1}{5}x_{k-3}+%
\frac{h}{45}\left( 26\,f_{k}-5\,f_{k-1}+4\,f_{k-2}-f_{k-3}\right) $ \\
&  \\ \hline
&  \\
$4$ & $x_{k}=2x_{k-1}-\frac{8}{5}x_{k-2}+\frac{26}{35}x_{k-3}-\frac{1}{7}%
x_{k-4}+\frac{h}{12600}(6463\,f_{k}-2092f_{k-1}$ \\
&  \\
& $\,\,\,\,\,\,\,\,\,\,\,+2298f_{k-2}-1132f_{k-3}+223f_{k-4}),$ \\
&  \\ \hline
&  \\
$5$ & $x_{k}=\frac{7}{3}x_{k-1}-\frac{38}{15}x_{k-2}+\frac{62}{35}x_{k-3}-%
\frac{43}{63}x_{{k-4}}+\frac{1}{9}x_{k-5}+\frac{h}{14175}%
(6669\,f_{k} $ \\
&  \\
& $\,\,\,\,\,\,\,\,\,\,\,-3122\,f_{k-1}+4358\,f_{k-2}-3192\,\,f_{k-3}+1253\,%
\,f_{k-4}-206\,f_{k-5}).$ \\
&  \\ \hline\hline
\end{tabular}
\caption{LIL algorithms.}
\end{center}
\end{table}

The study was achieved up to \ $m=8,$ \ but in this paper for the sake of
simplicity we considered only \ $m\in \{1,2,3,4,5\}$. For \ $m=1$ the LIL
method is equivalent to the backward Euler method.

The LIL method is an implicit method due to the presence of the term
$f_{k}$ in the right hand side which depends on $x_{k}$. Therefore
additional computations are necessary in order to calculate $f_{k}$.
In this purpose we approximate $x_{k-1} \in V_{k}$  (see
(\ref{2.1}))

\begin{equation*}
x_{k-1}\thickapprox x_{k}-\frac{h}{1!}x_{k}^{\prime }+\frac{h^{2}}{2!}%
x_{k}^{\prime \prime }-...+\left( -1\right) ^{m}\frac{h^{m}}{m!}%
x_{k}^{\left( m\right) }.
\end{equation*}
Using for derivatives the relations (2.3) one obtains
\begin{equation*}
x_{k-1}\thickapprox x_{k}-\frac{1}{1!}\sum\limits_{i=0}^{m}\delta
_{1\,i}x_{k-i}+\frac{1}{2!}\sum\limits_{i=0}^{m}\delta _{2\,i}x_{k-i}-...+%
\frac{\left( -1\right) ^{m}}{m!}\sum\limits_{i=0}^{m}\delta _{m\,i}x_{k-i}\,,
\end{equation*}
\noindent wherefrom we have
\begin{equation}
x_{k}\thickapprox \sum\limits_{i=1}^{m}\varepsilon
_{m\,i}x_{k-i},\quad i=1,2,...,m,\quad m>1.\,  \label{2.11}
\end{equation}
The coefficients \ $\varepsilon _{m\,i}$ \ are given in Table
9/Appendix.

\noindent Using (\ref{2.11}), \ $f_{k}$ \ becomes
\begin{equation*}
f_{k}=f\,\left( t_{k},\sum\limits_{i=1}^{m}\varepsilon _{m\,i}x_{k-i}\right)
.
\end{equation*}

The relation (\ref{2.11}) represents an extrapolation formula (predictor
phase) for \ $x_{k}$ \ and can be used to approximate the solution, but
without an acceptable accuracy, while (\ref{2.10}) is the corrector phase.

\noindent Because (\ref{2.10}) is a multi-step relation, a starting
method (for example the standard Runge-Kutta method) is necessary in
order to calculate the \ $m$ \ first start values: \
$x_{-1},x_{-2},...,x_{-m}$.

\section{The convergence}

\indent

The convergence is analyzed using the Dahlquist theory which states that a
numerical method is convergent\footnote{%
The ''convergence'' means here ''uniform convergence'' on an interval for
any \thinspace $C^{m}\,\,$smooth function $\,f$\thinspace .} if it is
consistent and stable (see [2], [4] or [7-8]). In this purpose let us
consider the LIL method (\ref{2.10}) in the usual form
\begin{equation}
\sigma _{10}x_{k}+\sigma _{11}x_{k-1}+...+\sigma _{1m}x_{k-m}=\sigma
_{00}\,f_{k}+\sigma _{01}\,f_{k-1}+...+\sigma _{0m}\,f_{k-m},
\label{3.1}
\end{equation}
\noindent with the characteristic polynomials
\begin{equation}
\alpha _{m}(s)=\sum\limits_{i=0}^{m}\sigma _{1\,i}\,s^{m-i},\quad
\beta _{m}\left( s\right) =\sum\limits_{i=0}^{m}\sigma
_{0\,i}\,s^{m-i},\,\,m\in \{1,2,3,4,5\}.  \label{3.2}
\end{equation}

\subsection{Consistency and errors}

\indent

Following the Dahlquist theory, the LIL method is consistent because its
characteristic polynomials (\ref{3.2}) satisfy \ $\alpha _{m}(1)=0$ \ and \ $%
\alpha _{m}^{\prime }(1)=-\beta _{m}\left( 1\right) $ \ for \ $m\in
\{1,2,3,4,5\}.$ \ As it is known, the order of a linear multi-step method is
$r$ \ if, and only if, \ $r$ $\ $of\ the following coefficients
\begin{equation*}
C_{j}=\sum\limits_{i=0}^{m}\sigma
_{1\,i}\,\,i\,^{j}+j\sum\limits_{i=0}^{m}\sigma _{0\,i}i^{\,j-1},\quad
j=1,2,...,r,
\end{equation*}
\noindent vanish.

\noindent Note that above the convention \ $0^{0}=1$ \ was used. The
values of \ $C$ \ for LIL method are given in Table 2.

\label{Table 2}
\begin{table}
\begin{center}
\begin{tabular}[t]{||c|c|c|c|c|c|c|c|c||}
\hline\hline
$m$ & $C_{1}$ & $C_{2}$ & $C_{3}$ & $C_{4}$ & $C_{5}$ & $C_{6}$ & $C_{7}$ & $%
\epsilon _{t}$ \\ \hline
{\small 1} & {\small 0} & {\small 0} & {\small -0.5} &  &  &  &  & $O(h^{2})$
\\ \hline
{\small 2} & {\small 0} & {\small 0} & {\small 0} & {\small -0.04} &  &  &
& $O(h^{3})$ \\ \hline
{\small 3} & {\small 0} & {\small 0} & {\small 0} & {\small 0} & {\small %
-0.313} &  &  & $O(h^{4})$ \\ \hline
{\small 4} & {\small 0} & {\small 0} & {\small 0} & {\small 0} & {\small 0}
& {\small -1.37} &  & $O(h^{5})$ \\ \hline
{\small 5} & {\small 0} & {\small 0} & {\small 0} & {\small 0} & {\small 0}
& {\small 0} & {\small -177.184} & $O(h^{6})$ \\ \hline\hline
\end{tabular}
\caption{C coefficients.}
\end{center}
\end{table}

\noindent From Table 2 one can deduce that the LIL order (the
largest \ $r$ \ for which \ $C$ is null) is \ $m+1.$ The local
truncation error \ $\epsilon _{t}$ \ is, for a given \ $m,$ \ of
order \ $m+1\,$(see e.g. [7]).

Comparatively, the local truncation error for the standard (4th-order)
Runge-Kutta algorithm is of order 4, and for the multi-step algorithms
Adams-Moulton and Gear are of order \ $m+1$, the same as for LIL algorithm.

The global truncation error (the accumulation of the local
truncation errors) per unit time is \ $\overline{\epsilon
_{t}}=\epsilon _{t}/h$. Hence the global truncation error per unit
time is of $\ m$ \ order.

\subsection{Stability}

\indent

LIL is stable if all solutions of the following difference equations

\begin{equation}
\alpha _{m}(s)=0,\,\,m\in \{1,2,3,4,5\},  \label{3.3}
\end{equation}

\noindent are bounded. A\ necessary and sufficient condition for stability
is that all zeros \ $s_{k}$\thinspace $,k=1,2,...,m$\ \ of \ $\alpha _{m}$ \
satisfy \ $\left| \,s_{k}\right| \leq 1$ \ and that zeros with \ $\left|
\,s_{k}\right| =1$ \ be simple. \ It is easy to see that $\,\alpha
_{1}(s)=s-1$ \ and\ for \ $m\geq 2,$\ $\,\,\alpha _{m}(s)=(s-1)\gamma
_{m-1}(s)$ \ (Table 3) with the zeros, numerically found for \ $%
m=3,4,5$, given in Table 4.

\label{Table 3}
\begin{table}
\begin{center}
\begin{tabular}{||l|l||}
\hline
$m=2$ & $\gamma _{1}(s)=(3\,s-1),$ \\ \hline
$m=3$ & $\gamma _{2}(s)=(15\,s^{2}-10\,s+3),$ \\ \hline
$m=4$ & $\gamma _{3}(s)=(35\,s^{3}-35\,s^{2}+21\,s-5),$ \\ \hline
$m=5$ & $\gamma _{4}(s)=(315\,s^{4}-420\,s^{3}+378\,s^{2}-180\,s+35).$ \\
\hline\hline
\end{tabular}
\caption{The polynomials $\gamma _{m-1}$.}
\end{center}
\end{table}

\label{Table 4}
\begin{table}
\begin{center}
\begin{tabular}{||c|c|c|c|c|c||}
\hline\hline
& $s_{1}$ & $s_{2}$ & $s_{3}$ & $s_{4}$ & $s_{5}$ \\ \hline
$m=2$ & $1$ & $0.33$ & {\small -} & {\small -} & {\small -} \\ \hline
$m=3$ & $1$ & $0.33+i\,0.30$ & $0.33-i\,0.30$ & {\small -} & {\small -} \\
\hline
$m=4$ & $1$ & $0.40$ & $0.30+i\,0.52$ & $0.30-i\,0.52$ & {\small -} \\ \hline
$m=5$ & $1$ & $0.40+i\,0.17$ & $0.40-i\,0.17$ & $0.26+i\,0.72$ & $%
0.26-i\,0.72$ \\ \hline\hline
\end{tabular}
\caption{The zeros of the characteristic equation $(3.3).$}
\end{center}
\end{table}

\noindent Hence the LIL method is stable and therefore we have the following
result

\begin{theorem}
The LIL method for to the initial value problem \emph{(\ref{1.1})} is
convergent for all \ $m\in \{1,2,3,4,5\}.$.

\begin{proof}
Because LIL is consistent and stable, following the Dahlquist theory, it is
convergent.
\end{proof}
\end{theorem}

\section{The regions of time stability}

\indent

An integration method may have low round-off error and low truncation error,
but be totally worthless because it is time unstable. The standard method
for testing the time (numerical) stability is to apply the integration
method to the first-order linear test equation

\begin{equation}
\overset{.}{x}=\lambda x,\quad x(0)=x_{0},  \label{4.1}
\end{equation}

\noindent where \ $x,$ $x_{0},\lambda $ \ may be complex. A method is \emph{%
time (numerically) stable} for specified values \ $\left( \lambda
,h\right) $ if it produces a bounded sequence \ $\{x_{n}\}$ \ when
applied to the test problem (\ref{4.1}) [7]. The set of the complex
values $\ z=\lambda h$ \ for which \ $\{x_{n}\}$ \ is bounded is
called the \emph{stability region} of the method. When an
integration method is applied to the system (\ref{4.1}) the result
is a linear, discrete-time system with a fixed point at the origin.
This means that the stability regions contain the half plan \
$Re(z)\leq 0.$ Therefore the stability of this fixed point
determines the time stability of the integration method.

\noindent Although this stability criterion guarantees that a method is
stable only when integrating a linear system, and not for nonlinear systems
it is an usual way to compare numerical performances for different
algorithms.

Following the theorem which states that a linear multi-step method is time
stable for a particular \ $z$ \ if and only if, the equation \ $\alpha
_{m}(\xi )=z\,\beta _{m}(\xi )$ \ has the following properties: all roots
satisfy \ $|\,\xi \,|\leq 1,$ \ and all roots with \ $|\,\xi \,|=1$ \ are
simple (see e.g. [8]), the proof of the time stability of LIL method follows
from convergence study.

In order to draw the stability regions let us define

\begin{equation*}
P_{m}(\xi ):=\alpha _{m}(\xi )-z\,\beta _{m}(\xi ),
\end{equation*}

Then, a linear multi-step method has the stability region \ $S$ , \ the set
of all points \ $z\in \mathbb{C}$ \ such that all the roots of \ $P_{m}(\xi
)=0$ \ lie inside or on the unit circle and those on the unit circle are
simple. Hence we obtain \ the equation $\ $

\begin{equation}
z=\frac{\alpha _{m}(\xi )}{\beta _{m}(\xi )},  \label{4.2}
\end{equation}

\noindent which has to be solved for any given \ $z\in \mathbb{C}$ . \ But
instead of solving (\ref{4.2}) for given \ $z$\thinspace , we can give\ $\xi
=e^{i\,\theta }$ \ with \ $\left| \,\xi \right| =1$ and plot\vspace{-0.09in}

\begin{equation}
z=\frac{\alpha _{m}(e^{i\,\theta })}{\beta _{m}(e^{i\,\theta })},
\label{4.3}
\end{equation}

\noindent for \ $\theta \in \left[ 0,\,2\pi \right] $ \ The set thus mapped
must contain \ $\partial \,S$ . The stability region of a numerical stable
algorithm has to contain the origin in his boundary.

\noindent In Figure 1 the stability regions for LIL algorithm for \thinspace
$m\in \{1,2,3,4,5\}\,$ are drawn. One can observe that LIL algorithm has,
for all \ $m,$ large (even unlimited) regions of stability, including the
entire left-half complex plane, typically for implicit algorithms$.$ \ The
time stability of LIL method is more efficient than that of other known
algorithms and is comparable with time stability of the Gear's algorithm
(see e.g. [5] where the stability regions were drawn for several known
algorithms).

\begin{figure}[ht]
\centering \subfigure[]{
\includegraphics[scale=0.53]{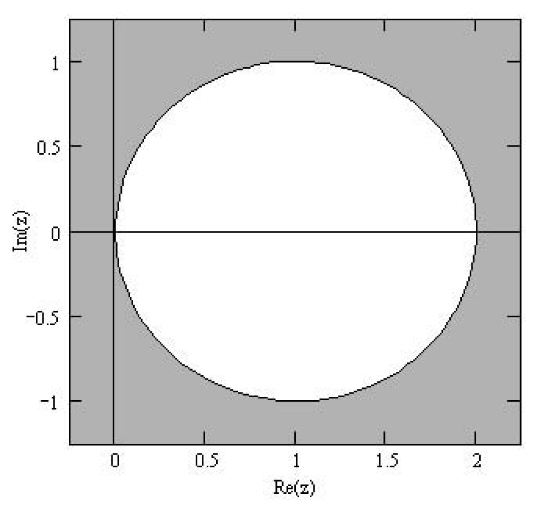}}
\subfigure[]{
\includegraphics[scale=0.5]{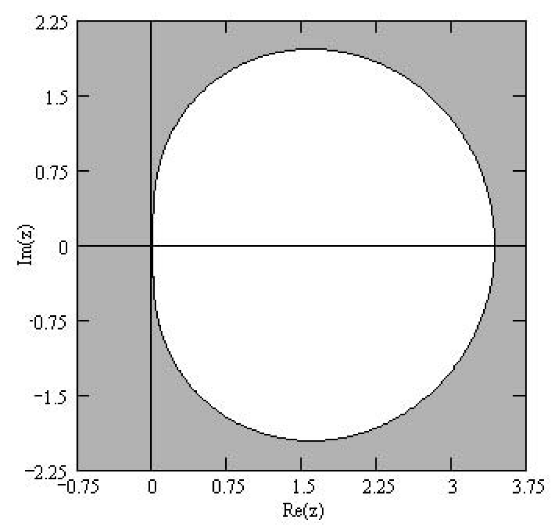}}

\centering \subfigure[]{
\includegraphics[scale=0.51]{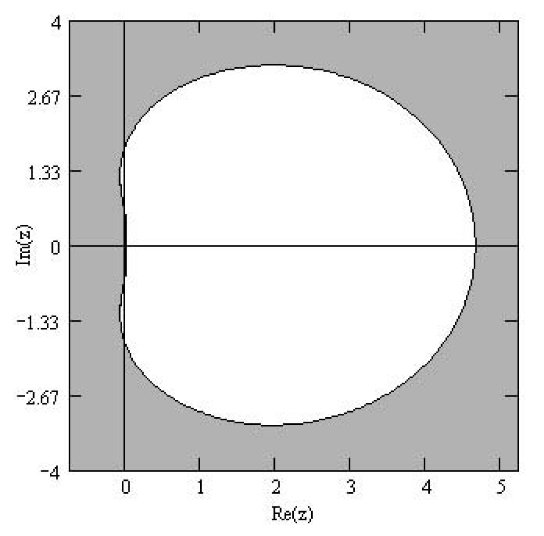}}
\subfigure[]{
\includegraphics[scale=0.52]{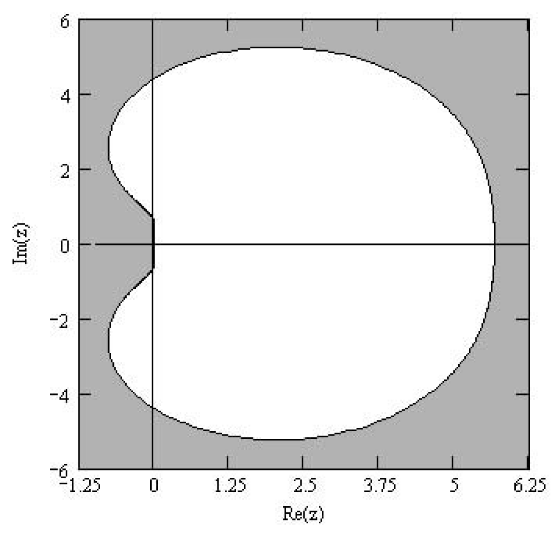}}

\begin{center}
\includegraphics[scale=0.5] {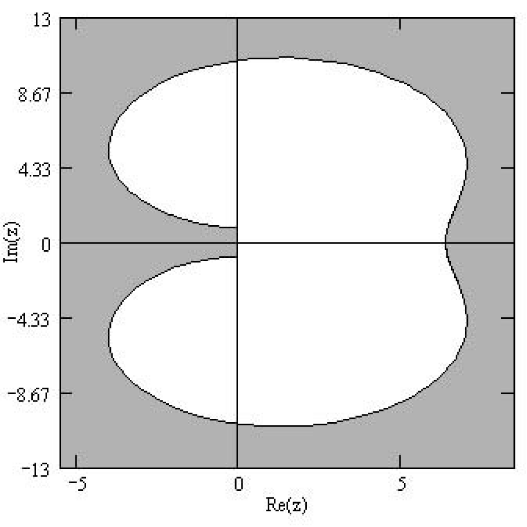}\label{fig1}
\caption{stability regions for $(m+1)$th LIL method: a) $m=1$; b)
$m=2$; c) $m=3;$ d) $m=4$; e) $m=5$.}
\end{center}
\end{figure}

\noindent Taking account of the fact that higher order is not always higher
accuracy, an acceptable compromise between the accuracy, time stability and
computational time was proved to be \ $m=3$.

\section{Applications}

\subsection{LIL versus standard methods}

The goal of this section is to compare the characteristics of few known
standard algorithms (the 4th\emph{-}order\emph{\ }$\,\,$methods:
Runge-Kutta, Gear, Adams-Moulton, the 3th\emph{-}order Adams-Bashforth
method and the Milne method) and 4th-order LIL method. For this purpose we
integrated two simple examples, with known analytical solutions: the
Bernoulli equation

\begin{equation*}
2\,t^{2}\overset{.}{x}(t)-4\,t\,x(t)-x^{2}(t)=0,\,
\end{equation*}

\noindent and

\begin{equation*}
\overset{.}{x\,}(t)=\cos (t).\,
\end{equation*}

\noindent The following values were calculated:

- the relative error \ $\varepsilon _{r}=\sum \left| x_{a}-x\right| /\sum
x_{a}\,,\ \ $where $x_{a}\,$\ is the analytical solution. The sum is taken
over the integration interval.

- the maximum absolute error: $\Delta =\underset{k}{\max }\left|
x_{a,k}-x_{k}\right| ,\,$\thinspace where $x_{a,k}$ \thinspace is the exact
solution in \thinspace $t_{k}.$

- the computation time $t$\footnote{$t$ is here only a relative value since
it depends on the used code (Turbo Pascal and using 64 bits), and the
computer processor (500 MHz ).}

\noindent The results are presented in Table 5 and 6.

\begin{table}[ht]
\begin{center}
\begin{tabular}{||c|c|c|c|c|c|c||}
\hline\hline (a) & R-K & Gear & A-M & A-B & Milne & LIL \\ \hline
$\varepsilon _{r}$ & {\small 1.9}$\cdot 10^{-4}$ & {\small
1.9}$\cdot 10^{-4} $ & {\small 1.1}$\cdot 10^{-7}$ & {\small
1.1}$\cdot 10^{-7}$ & {\small 1.4}$\cdot 10^{-7}$ & {\small
1.4}$\cdot 10^{-7}$ \\ \hline $\Delta $ & {\small 1.9}$\cdot
10^{-2}$ & {\small 2.0}$\cdot 10^{-2}$ &
{\small 1.2}$\cdot 10^{-5}$ & {\small 1.2}$\cdot 10^{-5}$ & {\small 1.8}$%
\cdot 10^{-5}$ & {\small 1.5}$\cdot 10^{-5}$ \\ \hline \emph{t}[s] &
{\small 0.16} & {\small 0.16} & {\small 0.16} & {\small 0.10} &
{\small 0.10} & {\small 0.16} \\ \hline\hline
\end{tabular}
\begin{tabular}{||c|c|c|c|c|c|c||}
\hline\hline (b) & R-K & Gear & A-M & A-B & Milne & LIL \\ \hline
$\varepsilon _{r}$ & {\small 3.8}$\cdot 10^{-5}$ & {\small
3.8}$\cdot 10^{-5} $ & {\small 2.3}$\cdot 10^{-10}$ & {\small
2.3}$\cdot 10^{-10}$ & {\small 2.8}$\cdot 10^{-10}$ & {\small
2.8}$\cdot 10^{-10}$ \\ \hline $\Delta $ & {\small 1.9}$\cdot
10^{-3}$ & {\small 2.0}$\cdot 10^{-3}$ &
{\small 1.2}$\cdot 10^{-8}$ & {\small 1.8}$\cdot 10^{-8}$ & {\small 1.8}$%
\cdot 10^{-8}$ & {\small 1.5}$\cdot 10^{-8}$ \\ \hline \emph{t}[s] &
{\small 0.82} & {\small 0.71} & {\small 0.82} & {\small 0.43} &
{\small 0.71} & {\small 0.87} \\ \hline\hline
\end{tabular}
\caption{ Bernoulli equation integrated with: a)
$h=0.01,~$\thinspace $t\in \lbrack 1,100]$; b) $h=0.001$, ~$t\in
[1,50].$}
\end{center}
\end{table}
\begin{table}
\begin{center}
\begin{tabular}{||c|c|c|c|c|c|c||}
\hline\hline $(a)$ & R-K & Gear & A-M & A-B & Milne & LIL \\ \hline
$\varepsilon _{r}$ & {\small 2.4}$\cdot 10^{-2}$ & {\small
4.9}$\cdot 10^{-2} $ & {\small 7.7}$\cdot 10^{-4}$ & {\small
4.0}$\cdot 10^{-3}$ & {\small 5.0}$\cdot 10^{-3}$ & {\small
5.0}$\cdot 10^{-3}$ \\ \hline $\Delta $ & {\small 3.7}$\cdot
10^{-2}$ & {\small 5.3}$\cdot 10^{-2}$ &
{\small 4.9}$\cdot 10^{-4}$ & {\small 2.6}$\cdot 10^{-3}$ & {\small 3.9}$%
\cdot 10^{-3}$ & {\small 3.3}$\cdot 10^{-3}$ \\ \hline
$\emph{t}$[s] & $0.0$ & {\small 0.0} & {\small 0.0} & {\small 0.0} & {\small %
0.0} & {\small 0.0} \\ \hline\hline
\end{tabular}
\begin{tabular}{||c|c|c|c|c|c|c||}
\hline\hline $(b)$ & R-K & Gear & A-M & A-B & Milne & LIL \\ \hline
$\varepsilon _{r}$ & $4.9\cdot 10^{-4}$ & $9.9\cdot 10^{-4}$ &
$3.9\cdot
10^{-7}$ & {\small 1.5}$\cdot 10^{-6}$ & {\small 1.9}$\cdot 10^{-6}$ & $%
1.9\cdot 10^{-6}$ \\ \hline $\Delta $ & $7.5\cdot 10^{-4}$ &
$1.0\cdot 10^{-3}$ & $2.4\cdot 10^{-7}$ & {\small 2.7}$\cdot
10^{-7}$ & {\small 1.5}$\cdot 10^{-6}$ & $1.2\cdot 10^{-6} $ \\
\hline $\emph{t}$[s] & $0.16$ & {\small 0.16} & {\small 0.16} &
{\small 0.16} & {\small 0.16} & {\small 0.16} \\ \hline\hline
\end{tabular}
\caption{ $\dot{x}(t)=\cos (t)$,  $t\in [0,2\pi]$ integrated with:
a) $h=0.05$; b) $h=0.001$.}
\end{center}
\end{table}

Comparing the results in Tables 5 and 6 one can deduce that LIL's
performances, for these two examples, are comparable to those of
performant methods like Gear, Adams-Moulton and Adams-Bashforth.

\subsection{Rabinovich-Fabrikant system}

\indent

The hard test was the integration of the Rabinovich-Fabrikant system.
Rabinovich and Fabrikant [6] studied the following dynamical system (named
the R-F model hereafter)

\begin{equation}
\begin{array}{l}
\overset{.}{x_{1}}=x_{2}(x_{3}-1+x_{1}^{2})+ax_{1}, \\[3pt]
\overset{.}{x_{2}}=x_{1}(3x_{3}+1-x_{1}^{2})+ax_{2}, \\[3pt]
\overset{.}{x_{3}}=-2x_{3}(b+x_{1}x_{2}),
\end{array}
\quad \qquad a,b\in \mathbb{R}.  \label{5.1}
\end{equation}

\begin{figure}[b]
\centering \subfigure[]{
\includegraphics[clip,width=0.6\textwidth]{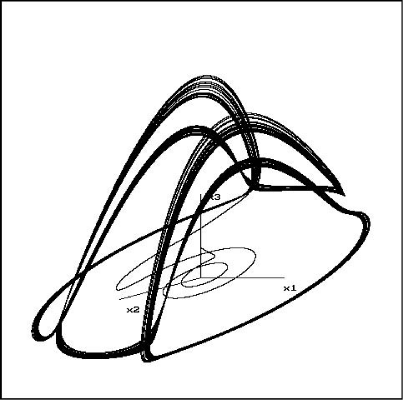}}
\end{figure}

\begin{figure}
\begin{center}
  \includegraphics[clip,width=0.6\textwidth] {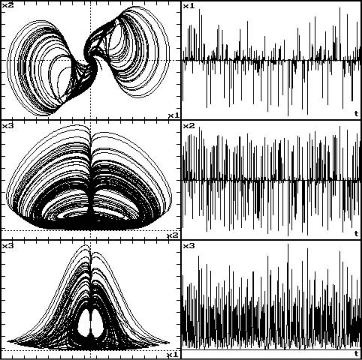}
  \caption{Two chaotic trajectories of R-F system: a)
  Three-dimensional phase portrait for $a=0.1$,~ $b=0.2876$; b)
  Plane phase portraits and time series for $a=-1$, ~ $b=-0.1$.}
\end{center}
\end{figure}

\begin{figure}
\centering \subfigure[]{
\includegraphics[clip,width=0.62\textwidth]{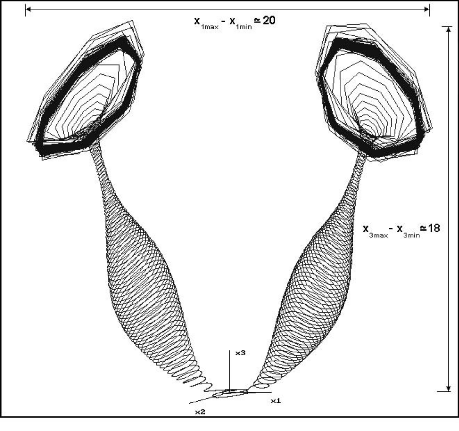}}
\end{figure}

\begin{figure}
\begin{center}
  \includegraphics[clip,width=0.62\textwidth] {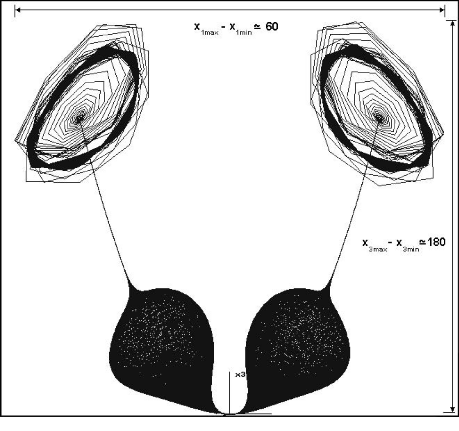}
  \caption{Two different sizes of the same attractor obtained with different
step-sizes: a) for \ $h=5\times 10^{-3},$ \ $x_{3\max }=35$ \ while
b) for \ $h=5\times 10^{-4},$ \ $x_{3\max }=350.$}
\end{center}
\end{figure}

\begin{figure}
\centering \subfigure[]{
\includegraphics[clip,width=0.55\textwidth]{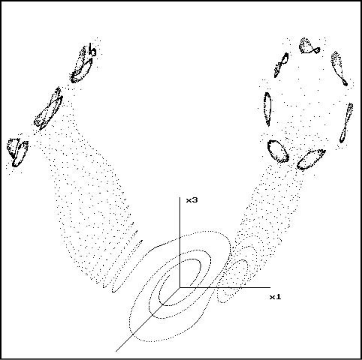}}
\end{figure}

\begin{figure}
\begin{center}
  \includegraphics[clip,width=0.55\textwidth] {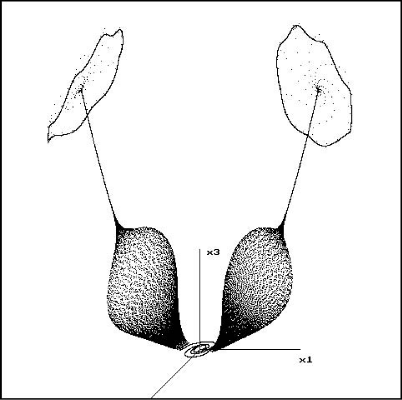}
  \caption{Two different attractors (plotted here by points), with the same
initial conditions and parameters values ($a=0.12,$ \ $b=0.05)$, but
with different step-size \ a) $h=0.05$ \ and \ b) $h=0.005.$}
\end{center}
\end{figure}

\begin{figure}
\centering \subfigure[]{
\includegraphics[clip,width=0.7\textwidth]{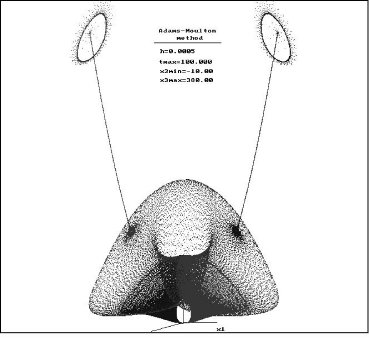}}
\centering \subfigure[]{
\includegraphics[clip,width=0.7\textwidth]{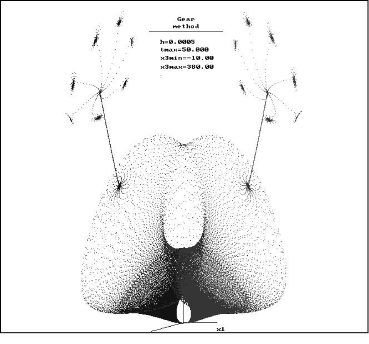}}
\end{figure}

\begin{figure}
\centering \subfigure[]{
\includegraphics[clip,width=0.6\textwidth]{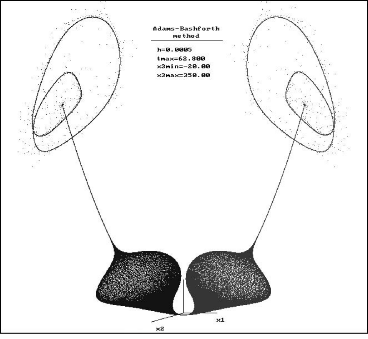}}
\end{figure}

\begin{figure}
\begin{center}
  \includegraphics[clip,width=0.6\textwidth] {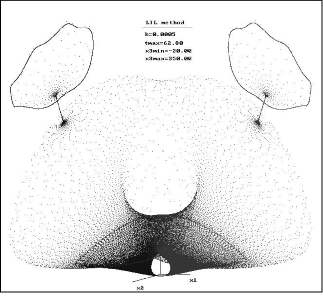}
  \caption{The case \ $a=0.3$ \ and \ $b=0.1$ \ integrated with: a) the 4%
\emph{th}-order Adams-Moulton ; b) the 4\emph{th}-order Gear algorithm; c) 3%
\emph{th} Adams-Bashforth algorithm; d) 4\emph{th}-LIL algorithm.}
\end{center}
\end{figure}

\noindent This system models the stochasticity arising from the modulation
instability in a non-equilibrium dissipative medium. Some qualitative
analysis and numerical dynamics have been reported in [6] and a carefully
re-examination together with many new and rich complex dynamics of the
model, that were mostly not reported before, are presented in [3]. The
chaotic R-F model proved to be a great challenge to the classical numerical
methods, most of them being not successful to study the complex dynamics of
this special model.

\noindent All computer test results and graphical plots in Figures 2-5 were
obtained with a special Turbo Pascal code which plots phase diagrams and
time series \ The code for LIL method may be obtained directly from the
author.

\noindent For $a<b$, the system is characterized by the appearance of
chaotic attractors in the phase space (see e.g. Figures 2).

\noindent It is well known that because of the sensitive dependence on
initial data, a chaotic system tends to amplify, often exponentially, tiny
initial errors. These kind of errors could be amplified to so large, that it
is almost impossible to draw mathematically rigorous conclusions based on
numerical simulations. A typical case can be seen from Figure 3, wherefrom
one deduces that the attractor's size along the $x_{3}$-axis increases
significantly as the step-size decreases. This problem has been noticed for
a long time, and has promoted a useful theory called ``shadowing,'' namely,
the existence of a true orbit nearby a numerically computed approximate
orbit [1]. \noindent We have also found that the strong dependence on the
step-size for R-F system, for certain values of $\ b$ and with the same
initial conditions, could produce totally different attractors (see Figures
4)

There are few special cases which proved to be a real challenge for the
numerical methods. As example for the case $\ a=0.3$ \ and \ $b=0.1$ (shown
in Figure 3), the 4th-order Runge-Kutta and Milne methods failed while only
the Gear and Adams-Moulton methods seem to give comparable results to those
obtained with LIL method; the attractors obtained with the 3th-order
Adams-Bashforth method are different to those obtained with Gear,
Adams-Moulton and LIL methods (Figures 5).

\section{Concluding remarks}

\indent

In this paper we present a linear implicit multi-step method, LIL, for ODEs
proving its convergence, too. The method could be considered as an
acceptable alternative to the classical algorithms for ODEs and can be
successfully used in practical applications. One of the advantages is that
in (\ref{2.5}) only the even order derivatives appear, this fact reducing
the truncation error and the computational time.

The algorithm seems to be stiffly-stable since it can integrate efficiently
and accurately enough dynamical systems like R-F which presents stiff
characteristics.

The implementation of adaptive step-size represents a task for a future
work. The basic approach would be applicable directly to variable step-size.

\section{Acknowledgments}

The author acknowledges Professor T. Colo\c{s}i, the promoter of this
method, for his continuous encouragement and discussions on this work.

\newpage

\begin{table}
\begin{center}
\textbf{Appendix}\\
\vspace{5mm}
\begin{tabular}{||c|c|c|c|c|c|c||}
\hline\hline
$i=2$ & $\delta _{20}$ & $\delta _{21}$ & $\delta _{22}$ & $\delta _{23}$ & $%
\delta _{24}$ & $\delta _{25}$ \\ \hline
$m=2$ & {\small 1} & {\small -2} & {\small 1} & $\cdots $ & $\cdots $ & $%
\cdots $ \\ \hline
$m=3$ & {\small 2} & {\small -5} & {\small 4} & {\small -1} & $\cdots $ & $%
\cdots $ \\ \hline
$m=4$ & {\small 35/12} & {\small -26/3} & {\small 19/2} & {\small -14/3} &
{\small 11/12} & $\cdots $ \\ \hline
$m=5$ & {\small 15/4} & {\small -77/6} & {\small 107/6} & {\small -13} &
{\small 61/12} & {\small -5/6} \\ \hline\hline
\end{tabular}
\begin{tabular}{||c|c|c|c|c|c|c||}
\hline\hline
$i=3$ & $\delta _{30}$ & $\delta _{31}$ & $\delta _{32}$ & $\delta _{33}$ & $%
\delta _{34}$ & $\delta _{35}$ \\ \hline
$m=3$ & {\small 1} & {\small -3} & {\small 3} & {\small -1} & $\cdots $ & $%
\cdots $ \\ \hline
$m=4$ & {\small 5/2} & {\small -9} & {\small 12} & {\small -7} & {\small 3/2}
& $\cdots $ \\ \hline
$m=5$ & {\small 17/4} & {\small -71/4} & {\small 59/2} & {\small -49/2} &
{\small 41/4} & {\small -7/4} \\ \hline\hline
\end{tabular}
\begin{tabular}{||c|c|c|c|c|c|c||}
\hline\hline
$i=4$ & $\delta _{40}$ & $\delta _{41}$ & $\delta _{42}$ & $\delta _{43}$ & $%
\delta _{44}$ & $\delta _{45}$ \\ \hline
$m=4$ & {\small 1} & {\small -4} & {\small 6} & {\small -4} & {\small 1} & $%
\cdots $ \\ \hline
$m=5$ & {\small 3} & {\small -14} & {\small 26} & {\small -24} & {\small 11}
& {\small -2} \\ \hline\hline
\end{tabular}

\begin{tabular}{||c|c|c|c|c|c|c||}
\hline\hline
$i=5$ & $\delta _{50}$ & $\delta _{51}$ & $\delta _{52}$ & $\delta _{53}$ & $%
\delta _{54}$ & $\delta _{55}$ \\ \hline
$m=5$ & {\small 1} & {\small -5} & {\small 10} & {\small -10} & {\small 5} &
{\small -1} \\ \hline\hline
\end{tabular}
\caption{$\delta$ coefficients.}\label{ap1} \vspace{8mm}
\begin{tabular}{||c|c|c|c|c|c||}
\hline\hline
$(a)$ & $m=1$ & $m=${\small 2} & $m=${\small 3} & $m=${\small \thinspace 4}
& $m=5$ \\ \hline
$\sigma _{00}$ & {\small 1} & {\small 25/24} & {\small 13/12} & {\small %
6463/5760} & {\small 741/640} \\ \hline
$\sigma _{01}$ & {\small 0} & {\small -1/12} & {\small -5/24} & {\small %
-523/1440} & {\small -1561/2880} \\ \hline
$\sigma _{02}$ & $\cdots $ & {\small 1/24} & {\small 1/6} & {\small 383/960}
& {\small 2179/2880} \\ \hline
$\sigma _{03}$ & $\cdots $ & $\cdots $ & {\small -1/24} & {\small -283/1440}
& {\small -133/240} \\ \hline
$\sigma _{04}$ & $\cdots $ & $\cdots $ & $\cdots $ & {\small 223/5760} &
{\small 1253/5760} \\ \hline
$\sigma _{05}$ & $\cdots $ & $\cdots $ & $\cdots $ & $\cdots $ & {\small %
-103/2880} \\ \hline\hline
\end{tabular}
\begin{tabular}{||c|c|c|c|c|c||}
\hline\hline
$(b)$ & $m=1$ & $m=${\small 2} & $m=${\small 3} & $m=${\small 4} & $m=$%
{\small 5} \\ \hline
$\sigma _{10}$ & {\small 1} & {\small 3/2} & {\small 15/8} & {\small 35/16}
& {\small 315/128} \\ \hline
$\sigma _{11}$ & {\small -1} & {\small -2} & {\small -25/8} & {\small -35/8}
& {\small -735/128} \\ \hline
$\sigma _{12}$ & $\cdots $ & {\small 1/2} & {\small 13/8} & {\small 7/2} &
{\small 399/64} \\ \hline
$\sigma _{13}$ & $\cdots $ & $\cdots $ & {\small -3/8} & {\small -13/8} &
{\small -279/64} \\ \hline
$\sigma _{14}$ & $\cdots $ & $\cdots $ & $\cdots $ & {\small 5/16} & {\small %
215/128} \\ \hline
$\sigma _{15}$ & $\cdots $ & $\cdots $ & $\cdots $ & $\cdots $ & {\small %
-35/128} \\ \hline\hline
\end{tabular}
\caption{$\sigma$ coefficients.}\label{ap2}\vspace{5mm}
\end{center}
\end{table}

\begin{table}
\begin{center}
\begin{tabular}{||c|c|c|c|c|c||}
\hline\hline
$m$ & $\varepsilon _{m1}$ & $\varepsilon _{m2}$ & $\varepsilon _{m3}$ & $%
\varepsilon _{m4}$ & $\varepsilon _{m5}$ \\ \hline
$2$ & {\small 2} & {\small -1} & $\cdots $ & $\cdots $ & $\cdots $ \\ \hline
$3$ & {\small 3} & {\small -3} & {\small 1} & $\cdots $ & $\cdots $ \\ \hline
$4$ & {\small 4} & {\small -6} & {\small 4} & {\small -1} & $\cdots $ \\
\hline
$5$ & {\small 5} & {\small -10} & {\small 10} & {\small -5} & {\small 1} \\
\hline\hline
\end{tabular}
\caption{ $\varepsilon $ coefficients.}\label{ap3}
\end{center}
\end{table}

\end{document}